\documentclass[%
 aip,
 amsmath,amssymb,
 reprint,%
]{revtex4-1}

\usepackage{graphicx}
\usepackage{dcolumn}
\usepackage{bm}

\usepackage[utf8]{inputenc}
\usepackage[T1]{fontenc}
\usepackage{mathptmx}
\usepackage{etoolbox}
\usepackage[version=4]{mhchem}
\usepackage{array}
\makeatletter
\def\@email#1#2{%
 \endgroup
 \patchcmd{\titleblock@produce}
  {\frontmatter@RRAPformat}
  {\frontmatter@RRAPformat{\produce@RRAP{*#1\href{mailto:#2}{#2}}}\frontmatter@RRAPformat}
  {}{}
}%
\begin{document}


\title{Development of a projectile charge state analyzer and 10 kV bipolar power supply for MeV energy ion $-$ atom/molecule collision experiments}
\author{Sandeep Bajrangi Bari}
\affiliation{Department of Physics, Indian Institute of Technology Kanpur, Kanpur - 208016, India.}
\author{Sahan Raghava Sykam}
\affiliation{Department of Physics, Indian Institute of Technology Kanpur, Kanpur - 208016, India.}
\author{Ranojit Das}
\affiliation{Department of Physics, Indian Institute of Technology Kanpur, Kanpur - 208016, India.}
\author{R. Tyagi}
\affiliation{Department of Physics, Indian Institute of Technology Kanpur, Kanpur - 208016, India.}
\author{A. H. Kelkar}
\affiliation{Department of Physics, Indian Institute of Technology Kanpur, Kanpur - 208016, India.}
\email{akelkar@iitk.ac.in}
\date{\today}

%

\date{\today}

\begin{abstract}
We have developed a post-collision projectile charge state analyzer (CSA) for detecting the charge state of the projectile ion following ion-atom/molecule collision. The design of the analyzer, based on electrostatic parallel plate deflector was simulated using SIMION ion optics package. We have also developed a 10 kV bipolar programmable power supply to bias the CSA electrodes. The CSA and the power supply, both, were tested in collision studies using MeV energy ion beam of proton and carbon ions at the 1.7 MV tandetron accelerator facility at IIT Kanpur.
\end{abstract}

\maketitle


\section{Introduction}

Interaction of ions with atomic or molecular targets results in reorganization of the target and projectile electronic structure. Fundamental processes such as ionization, electron capture and transfer ionization characterize the energy transfer processes. In depth investigation of these processes reveals the nature of Coulomb interaction between charged and neutral particles. Ionization and electron capture cross sections measured for various projectile - target combinations are also useful in allied fields such as radiation biology, plasma physics, astrochemistry etc.

Ion collisions with atoms or molecules where the projectile ion may undergo a change in its initial charge state (due to electron loss or capture) require measurement of the projectile ion charge state, post collision, in coincidence with the ionized target ions. In charge exchange collisions, the projectile undergoes a change in its charge state. This may be accompanied by target ionization where the target loses an electron in the continuum resulting in the process of transfer ionization. If the electron is captured from an inner shell, the vacancy created initiates a cascade of Auger - Meitner decay resulting in multiple ionization of the target. The filling of vacancy created by charge exchange may also result in emission of X-rays.

The charge state of the projectile ion, can be analyzed using electric fields and magnetic fields. In most cases the fields are static and uniform leading to a simple and easy to implement design. Various designs of projectile charge state analyzers based on parallel plate, and cylindrical or spherical sectors as well as quadrupole fields, magnetic sector analyzers, etc. are widely used \cite{sar1967cylindrical, kreckel2010simple, siddiki2022development,srivastav2022post, knudsen1981single}. The design considerations
depend on the energy of the ion beam and the required charge state separation. The electrostatic parallel plate deflector is a preferred choice due to its simple and low cost design for ion beam energies ranging rom few keV/q to tens of MeV/q. 

In this work we present the design and implementation of a parallel plate projectile charge state analyzer for MeV energy ion beams. We have also constructed a 10 kV bipolar power supply for the analyzer electrodes. The power supply has remote operation capability and can be easily integrated with Arduino microcontroller. The projectile charge state analyzer setup was integrated with a momentum spectrometer \cite{duley2022design} to study electron capture induced ionization and fragmentation of atomic and molecular targets. 

\section{Projectile charge state analyzer}

\begin{figure*}[tbp]
    \centering
    \includegraphics[width=\textwidth]{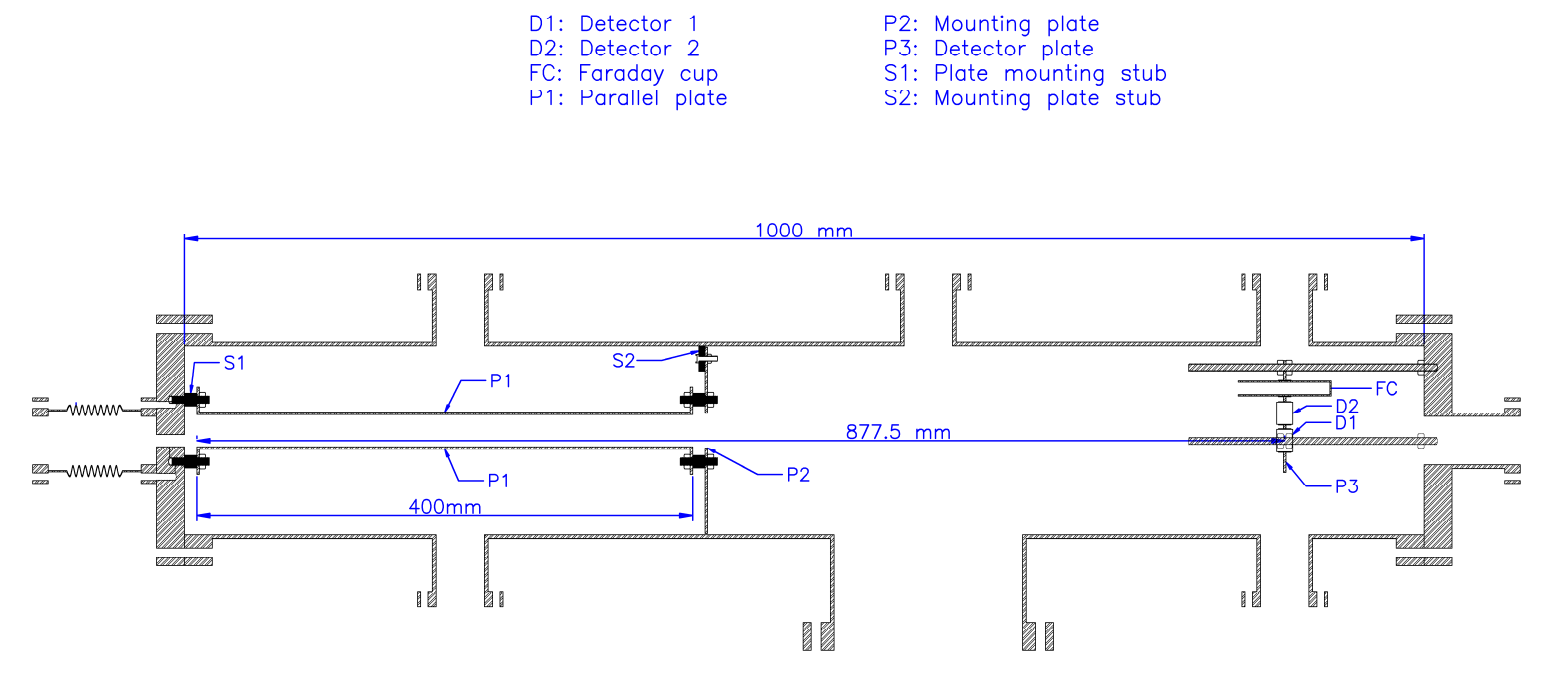}
    \caption{Schematic of the projectile charge state analyzer. The mounting stubs S1 are fabricated from Delrin\textregistered{ }and S2 from Teflon\textregistered{}. All other parts are fabricated from non-magnetic stainless steel. Detectors D1 and D2 are either CEMs or SBDs, chosen based on the energy of the incident ion beam and resolution of the time of flight spectrum. The FC and the detectors D1 and D2 are electrically isolated from the detector plate.}
    \label{fig1}
\end{figure*}

\begin{figure}[tbp]
    \centering
    \includegraphics[width=0.45\textwidth]{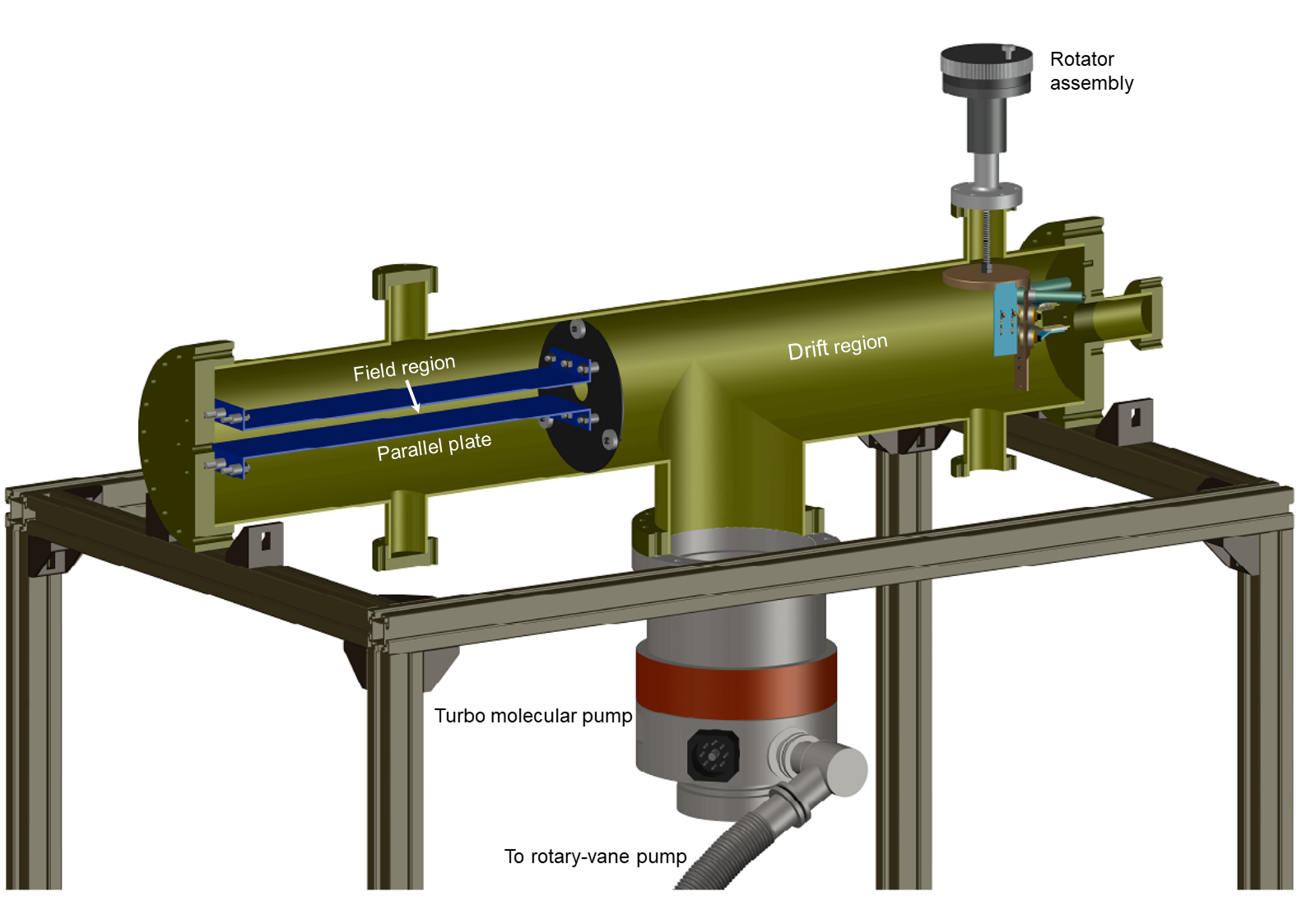}
    \caption{3D CAD assembly drawing of the CSA with rotatable detector assembly.}
    \label{fig2}
\end{figure}

\begin{figure}[tbp]
    \centering
    \includegraphics[width=0.45\textwidth]{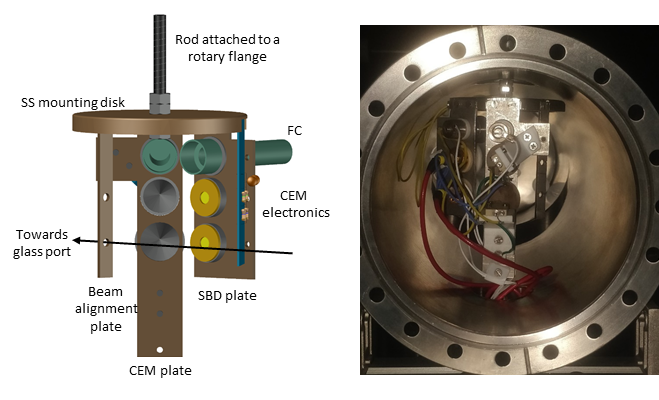}
    \caption{CAD rendering of the rotatable detector assembly (left) and photograph of the detector assembly mounted in the CSA chamber (right). Vertical positioning of the apertures in beam alignment plate coincides with the centres of the CEMs and SBDs in the CEM plate and SBD plate respectively. The rotary enables rotation of the mounting SS disk to bring the detector plates in line with the incident beam after alignment is done using the beam aliggnment plate.}
    \label{fig3}
\end{figure}

The projectile charge state analyzer has been designed to be used in conjunction with a recoil ion momentum spectrometer (RIMS) \cite{duley2022design} at the 1.7 MV Tandetron accelerator facility at IIT Kanpur. The present configuration allows identification of single and double electron capture to projectile ions with maximum energy of 5 MeV/q. Figure \ref{fig1} shows the CAD drawing of the projectile charge state analyzer (CSA) assembly. The entire assembly is placed in a 1 m long stainless steel (SS-304) vacuum chamber (diameter = 150 mm) connected to a 450 lt/s turbomolecular pump backed by a 10 m$^3$/h rotary vane backing pump. The vacuum in the CSA chamber is maintained at $\leq$ 10$^{-7}$ mbar. The CSA consists of two regions, an electric field region and a field-free drift region. The electric field region is formed by two rectangular metal (SS-304) plate electrodes forming a parallel plate deflector. Appropriate voltages of opposite polarity are applied via high vacuum electrical feedthroughs to the two plate electrodes to maintain the requisite electric field. The plates are mounted on a CF-100 flange with an aperture of 10 mm diameter for beam entry. The other end the plates are supported by a SS disc with a central hole (diameter = 28 mm) for the ion beam to pass through. The plate electrodes have a length of 400 mm, and width of 60 mm. Separation between the plates is 26 mm. The plates are mounted off-center with respect to the ion beam axis. The distance between the bottom plate and the geometric axis of the CSA vacuum chamber is 10 mm. The asymmetric mounting allows maximum utilization of the deflecting electrode gap region for beam bending. This leads to a lower deflecting voltage compared to a symmetric plate arrangement. The projectile beam traverses a field free drift region upon exiting the deflecting field. The drift region extends for 470 mm, where a rotatable detector assembly is mounted to detect the charge analyzed projectile ions.

The rotatable detector assembly consists of a set of three rectangular metal plates mounted perpendicular on the rim of a rotatable stainless steel disc. One of the plates is used to mount a pair of silicon surface barrier detectors (SSBDs) and the other one holds a pair of channel electron multipliers (CEMs). A Faraday cup for measuring the projectile beam current is also mounted on each detector plate. The center-to-center distance between the two detectors is 22 mm  and the Faraday cup is placed  at a distance of 20 mm from the center of the upper detector. The third metal plate connected to the assembly is used for projectile beam alignment. The beam alignment plate has two circular holes for aligning the projectile beam. The centers of the apertures lie at the same position as those of the detectors. The three rectangular plates are placed $\mathrm{40^o}$ apart. The SSBDs are most suitable for MeV energy heavy ion detection. These detectors are regularly used in Rutherford back-scattering experiments. However, the raw (and preamplifier) output pulse of an SSBD has a width of few micro seconds. Therefore the signal from SSBD leads to reduced time-of-flight resolution during the experiments. In comparison, the CEM provides a better time resolution as the output pulse is only a few nanoseconds wide. The rotatable detector mechanism allows in-situ switching between the detectors.

\begin{figure*}[tbp]
    \centering
    \includegraphics[width=\textwidth]{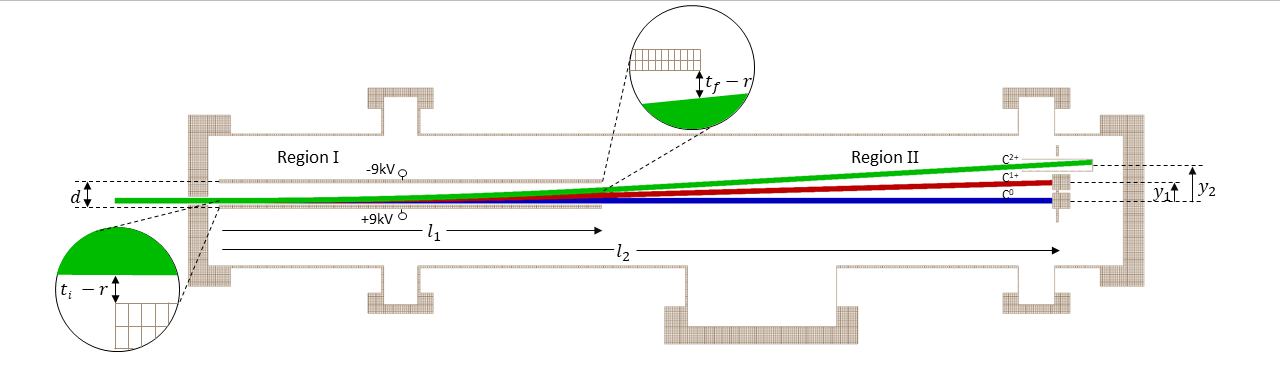}
    \caption{SIMION simulation of the CSA for an incident carbon beam with an energy of 4.5 MeV. The beam is a mixture of C$^{2+}$,  C$^{+}$ ions and  C neutral atoms with identical kinetic energy. Tolerances at the entry and exit of Region I are denoted as t$_i$ and t$_f$, respectively. Under the influence of the electric field applied between the deflecting plates, the composite beam separates into its constituent charge/neutral species.}
    \label{fig4}
\end{figure*}
\subsection{Simulation results}
SIMION 8.0 was used to optimize the electrode potential and geometry of the CSA. Fig.~\ref{fig4}
shows the simulated trajectories for an incident carbon beam with an
energy of 4.5 MeV. The beam is a mixture of C$^{2+}$, C$^+$ ions and neutral C atoms. Two parallel plates each of length $\mathrm{l_1}$ and width $\mathrm{w}$ 
were placed a distance $\mathrm{d}$ apart. A cylindrical carbon beam of energy $\mathrm{E}$ and radius $\mathrm{r}$, which is composed of charge states q = 0, 1 and 2, enters the region between the plates at an
axial distance of $\mathrm{t_i - r}$ mm from the lower plate. $\mathrm{t_i}$ is the tolerance at the entrance point of the beam in the region between the plates. The system is designed for detection of positively charged ions. 
A symmetric potential difference of $\mathrm{ \pm V}$ volts is applied between the plates to deflect the ion beams away from the axis. The neutral particles are not affected by this field and continue traveling along the original beam direction.  
After exiting the field region (region $\mathrm{I}$),
the beams further separate in the drift region (region $\mathrm{II}$ ). At the end of the drift region, the  charge separated ion beams are detected by a pair of detectors and a Faraday cup, in the detector plane at a distance $l_2$ from the the point of beam entrance. One of the detectors
is located on the original axis of the projectile beam for detecting the neutral particles. The detector for single capture and a Faraday cup for main projectile beam current measurement are located at a distance of $\mathrm{y_1}$ and $\mathrm{y_2}$ from the axis of the projectile beam.  Figure \ref{fig14} shows the trajectory of the $\mathrm{C^{2+}}$, $\mathrm{C^{+}}$, and $\mathrm{C^{0}}$ beams that separate from the incoming projectile beam. The ion beam separation depends only on the energy and charge state of the projectile ions. To optimize the design of the CSA, we had three main constraints. (i) The length of the CSA assembly was fixed at 1 m as per the availability of space in the 20$\mathrm{^o}$ beamline of the Tandetron accelerator. (ii) The energy of the projectile beam was taken as 4.5 MeV ($\mathrm{q = 2+}$) and its size was taken as 6 mm diameter. (iii) The maximum voltage applied to the plate electrodes was kept at 10 kV. With these constraints, the CSA dimensions and detector positions were optimized using SIMION 8.0. In table \ref{table1} we have listed the optimized values. In order to verify the optimized values, a cylindrical beam, consisting of $\mathrm{C^0}$, $\mathrm{C^+}$ and $\mathrm{C^{2+}}$, of radius 3 mm was passed at an axial distance of 5 mm from the lower plate. The plate voltages were varied to obtain the hit position of 42 mm in the detector plane for the doubly charged ion $\mathrm{C^{2+}}$. The hit positions for the singly charged ion $\mathrm{C^+}$ (single capture) and the undeflected neutral atom $\mathrm{C^0}$ (double capture) were also recorded. The simulation was repeated for different ion beam energies and the plate voltages were noted for the same hit positions of the ions in the detector plane. These values are recorded in table \ref{table2} for beam energies ranging from 1 MeV to 4.5 MeV.

\begin{table}[tbp]
\begin{ruledtabular}
\begin{tabular}{cccccc}
  $\mathrm{t_i}$ & $\mathrm{t_f}$ & $\mathrm{l_1}$ & $\mathrm{l_2}$ & $\mathrm{y_1}$ & $\mathrm{y_2}$\\
  \hline
  2 & 5 & 400 & 870 & 25 & 45\\
\end{tabular}
\end{ruledtabular}
\caption{Geometrical parameters of the CSA optimized in SIMION as shown in figure \ref{fig4} (all dimensions in mm).}
\label{table1}
\end{table}

\begin{table}[tbp]
\begin{ruledtabular}
\begin{tabular}{cc}
  Projectile energy (MeV/q) & CSA electrode voltage (kV) \\ 
  \hline
  1.0 & $\pm$ 2\\
  1.5 & $\pm$ 3\\
  2.0 & $\pm$ 4\\
  2.5 & $\pm$ 5\\
  3.0 & $\pm$ 6\\
  3.5 & $\pm$ 7\\
  4.0 & $\pm$ 8\\
  4.5 & $\pm$ 9\\
\end{tabular}
\end{ruledtabular}
\caption{
Simulated values of potential difference between the CSA electrodes for deflection of singly charged ions by $1.45^{\circ}$ 
detected at 871 mm and deflection of doubly charged ions by $2.86^{\circ}$ detected at 840 mm.
}
\label{table2}
\end{table}

\subsection{Calibration of the CSA}

\begin{figure}[tbp]
    \includegraphics[width=0.5\textwidth]{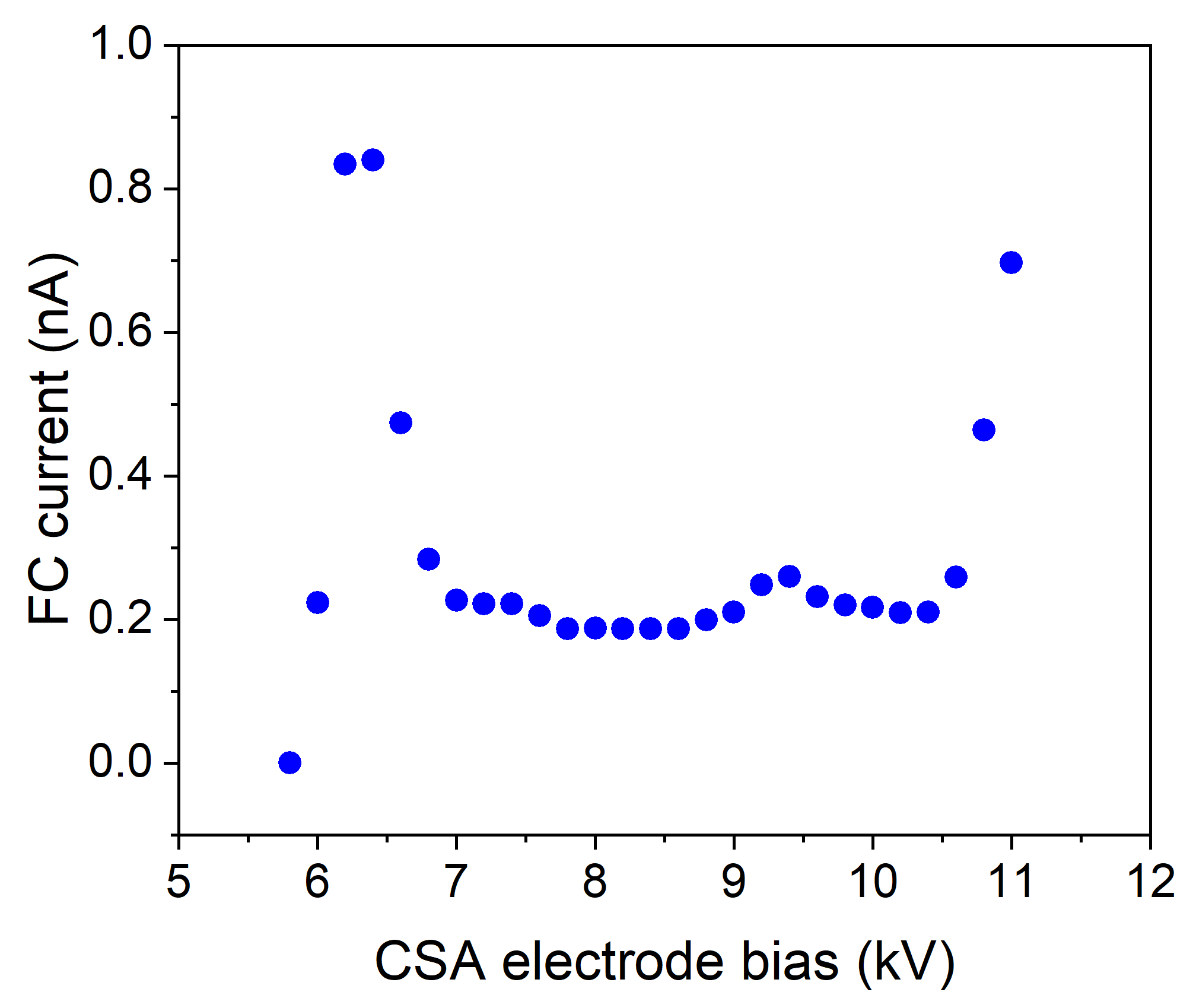} 
    \caption{Measured beam current in FC as a function of applied potential difference between the electrodes. Energy of the incident projectile beam is 1.0 MeV}
    \label{fig5}
\end{figure}

\begin{figure}[tbp]
   \includegraphics[width=0.5\textwidth]{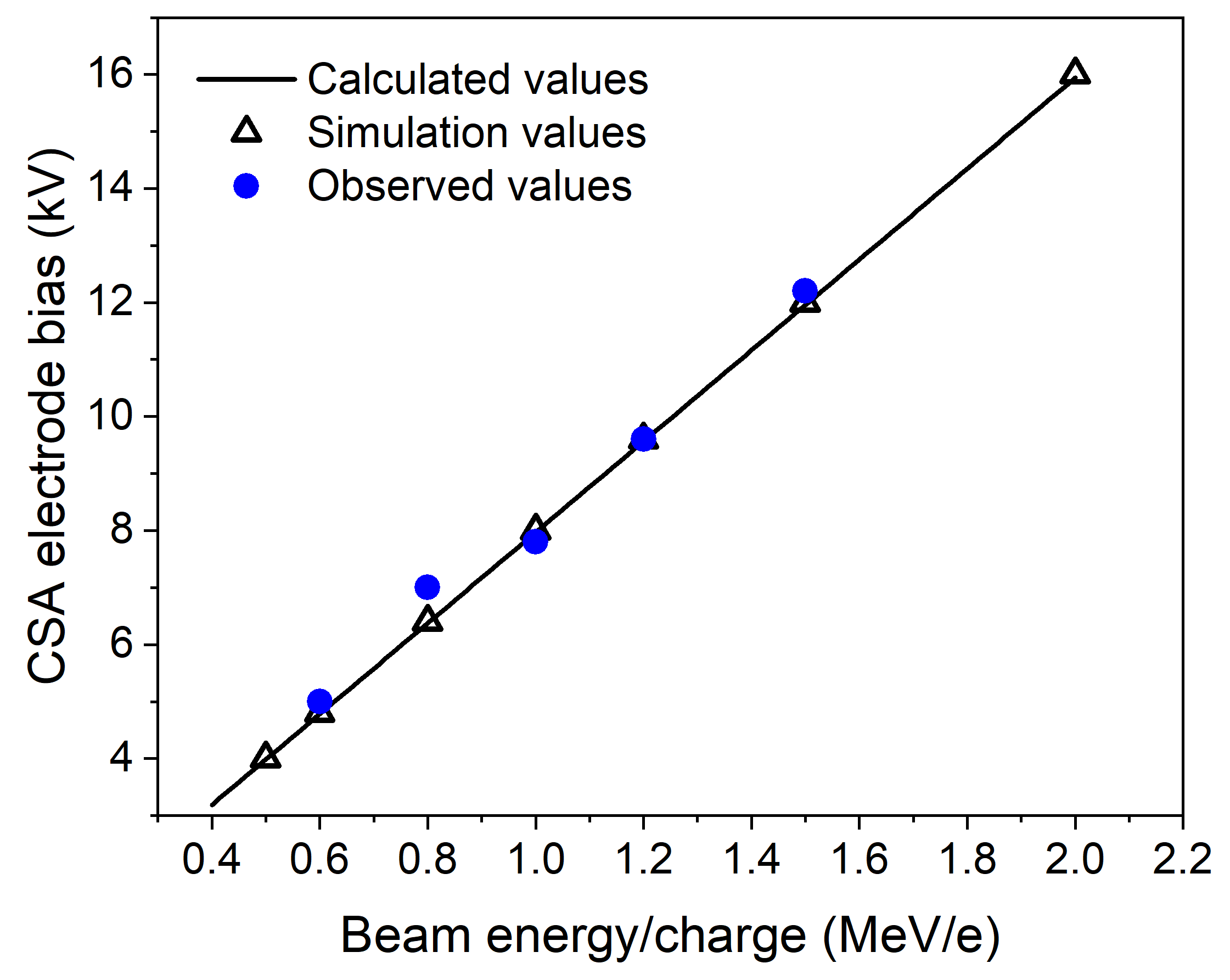}
    \caption{Projectile ion beam energy per unit charge vs CSA electrode potential. The solid blue circles are the measured values and open black triangles are values obtained though simulations. The solid black line represents analytical relation between the electrode potential and ion beam energy for a fixed deflection.}
    \label{fig6}
\end{figure}

Performance of the CSA was tested using 500 keV to 3 MeV proton beam obtained from Duoplasmatron source of the 1.7 MV Tandetron accelerator facility at IIT Kanpur. The CSA was calibrated in a manner similar to the simulations performed in SIMION. For a given ion beam energy, the potential difference between the CSA plates was varied until the beam hits symmetrically at the centre of the Faraday cup fixed at a distance of 42 mm from the beamline axis. This was repeated for $\mathrm{H^+}$ beams of different energies. For 1 MeV H$^+$ beam passing through CSA, the potential difference between the CSA plates was varied from 6 kV to 11 kV and the beam current in FC was noted. Figure \ref{fig5} shows the intensity profile of the $\mathrm{H^+}$ ion beam measured on the Faraday cup as a function of the CSA electrode voltage. For voltages below 5.5 kV, the current measured in FC is zero. With increasing electrode voltage, the beam intensity rises steeply till 6.4 kV. The steep rise is due to the beam falling on the outer surface of the FC. Energetic ion beam hitting a metallic surface results in secondary electron emission. Loss of these secondary electrons from the surface of the FC is recorded as an increase in the positive ion current. Therefore, the current measured by FC has two components. One due to the beam falling on the FC and the other due to secondary electron emission. For electrode potential above 6.4 kV, the current starts to decrease and reaches a minimum at 7.8 kV and stays constant over the range 7.8 to 8.6 kV. In this range, the ion beam falls completely within the FC. Due to the length of the FC, the secondary electrons emitted from the inner surface of the FC are absorbed back into the FC. Therefore, the current shown by FC in this range is the actual projectile beam current. We have selected the central value of the range 7.8 to 8.6 kV as the bias voltage required to obtain the detector hit position in the detector plane (for 1 MeV singly charged ions). At higher voltages beyond 8.7 kV the current starts rising again. This is due to beam falling at the rim of FC so that the secondary electrons emitted at the FC surface are not completely absorbed back resulting in a value higher than the incident beam current. At even higher voltages, the measured beam current quickly falls down to zero as there is no ion falling on the FC.  In a similar manner, appropriate bias voltages were obtained for the same hit position on the FC for a range of incident projectile energies for $\mathrm{H^+}$, $\mathrm{C^+}$ and $\mathrm{C^{2+}}$ projectile ion beams

Figure \ref{fig6} shows the ion beam energy vs electrode potential calibration plot for the CSA. In the figure, we have also plotted the calibration data obtained from the simulations. The agreement between the measured and simulated data is excellent. 

\section{10 kV bipolar power supply}

\begin{figure*}[tbp]
    \centering
    \begin{tabular}{c}
           \includegraphics[width=0.86\textwidth]{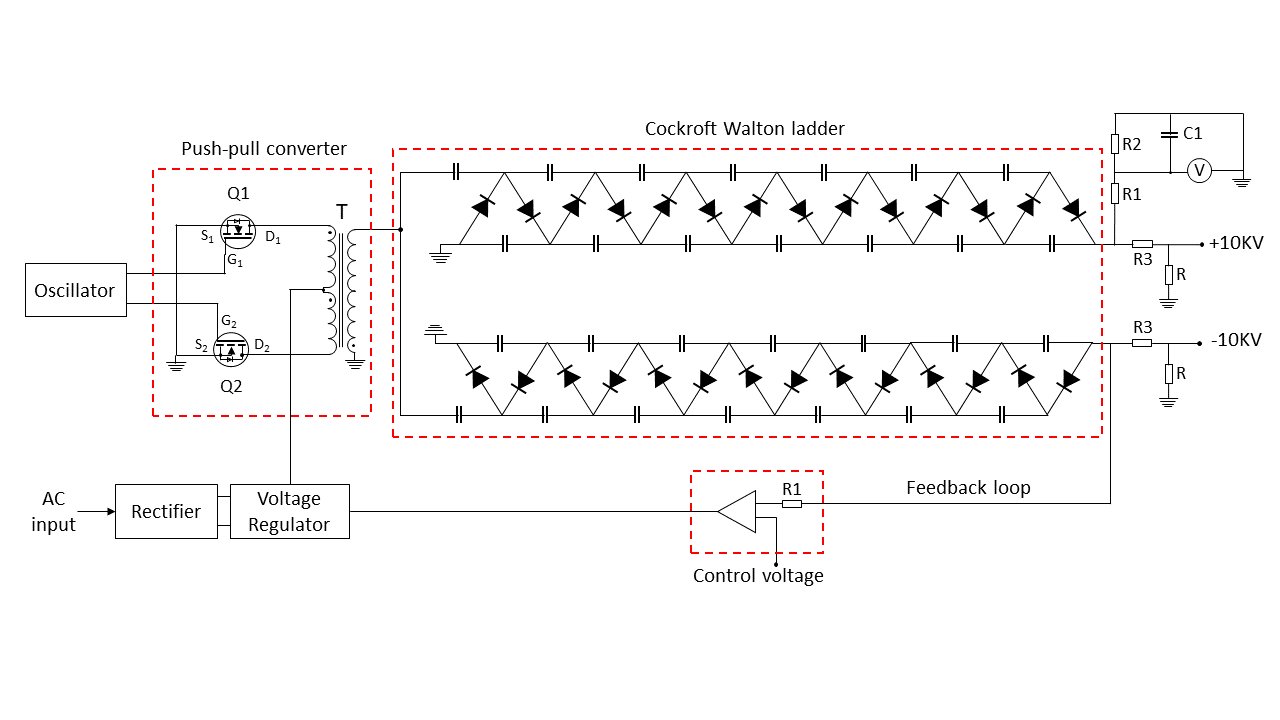} \\
         \includegraphics[width=0.65\textwidth]{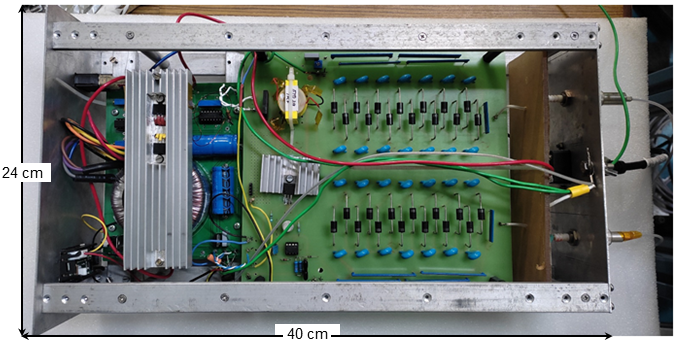}
    \end{tabular}
    \caption{Circuit diagram (upper figure) and photograph (lower figure) of the bipolar CCW power supply. Q$_1$ and Q$_2$ are MOSFETs driving the push-pull converter. G,S and D represent the gate, source and drain of each of the MOSFETs respectively.}
    \label{fig7}
\end{figure*}

A high voltage dc power supply was designed and assembled in-house for biasing the CSA electrodes. The bipolar dc power supply developed in the lab is based on a Cockcroft–Walton (CCW) voltage multiplier architecture. This approach eliminates the need for bulky transformers that typically require large iron cores and extensive insulation, there by enhancing compactness and efficiency. Figure \ref{fig7} shows the circuit diagram of the power supply. Specifications of the electrical components are listed in table \ref{table3}. In the following, we shall discuss the operating characteristics of the power supply in brief.

\begin{table}[h!]
\centering
\begin{ruledtabular}
\begin{tabular}{ll}

Components & Specification/Part number \\ \hline

R & 150 M$\Omega$ \\

R$_1$ & 150 k$\Omega$ \\

R$_2$ & 1 M$\Omega$ \\

C$_1$ & 100 nF \\

Oscillator & SG3525 \\
\hline
Transformer & turn ratio: 1:54 \\
core dimension: E30/15/7 &  \\
core material: N27 & \\
\hline
Rectifier & 800 V, 6 A, GBU6K \\

Voltage regulator & XL4015 \\

CCW Capacitor & 3.2 nF \\

CCW Diode & 15 kV, 0.1 A \\

Q$_1$, Q$_2$ & 100 V, 9.2 A \\

\end{tabular}
\end{ruledtabular}
\caption{Specifications of the various electrical components used in the circuit of the bipolar 10 kV power supply (see fig. \ref{fig7}).}
\label{table3}
\end{table}

\begin{figure}[tbp]
     \includegraphics[width=0.45\textwidth]{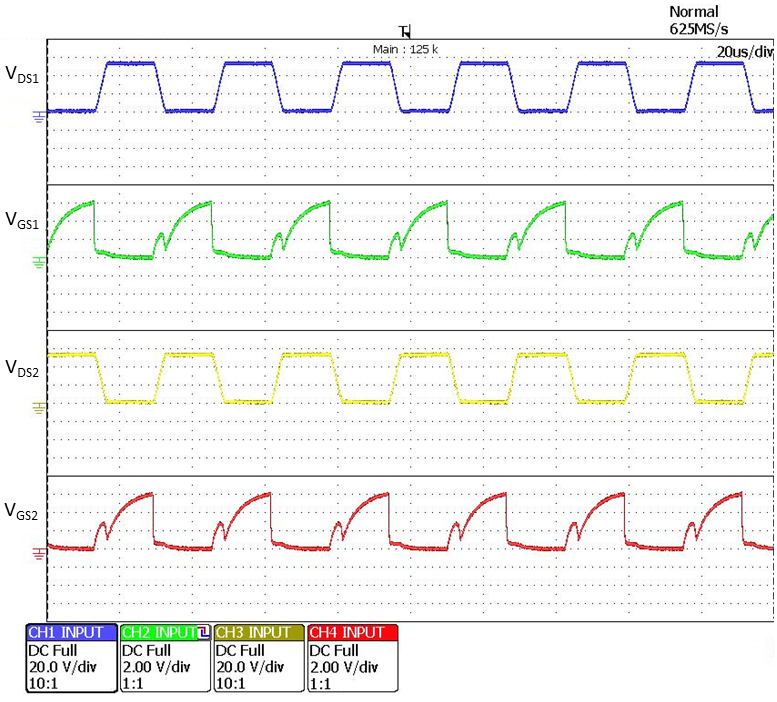}
    \caption{The input and output waveforms of the transformer as measured with the oscilloscope (Yokogawa Model: DL9140). The gate-source voltages $\mathrm{V_{GS1}}$ and $\mathrm{V_{GS2}}$ are input waveforms from the oscillator to the MOSFETs. The two waveforms are phase-shifted by $\pi$ radians. $\mathrm{V_{DS1}}$ and $\mathrm{V_{DS2}}$ are the waveforms at the drains of the two MOSFETs $\mathrm{Q_1}$ and $\mathrm{Q_2}$.}
    \label{fig8}
\end{figure}

\begin{figure}[tbp]
    \includegraphics[width=0.45\textwidth]{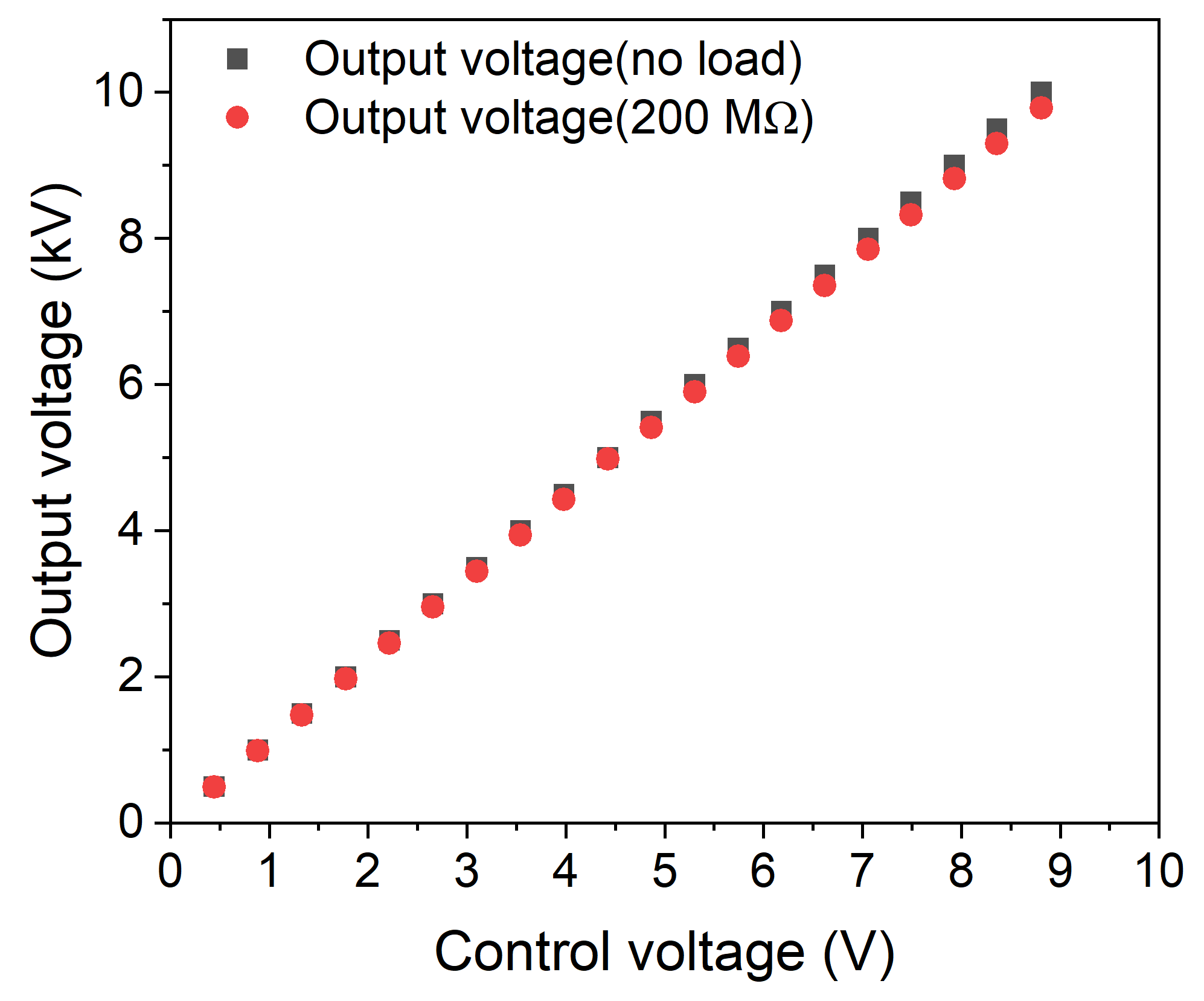} 
    \caption{High voltage output of the CCW power supply as a function of applied control voltage. Solid black circles show output voltage under no load (open) condition. Solid red circles correspond to a load of $\mathrm{200 M\Omega}$ connected between power supply output and ground terminal.}
    \label{fig9}
\end{figure}

\begin{figure}[tbp]
    \includegraphics[width=0.45\textwidth]{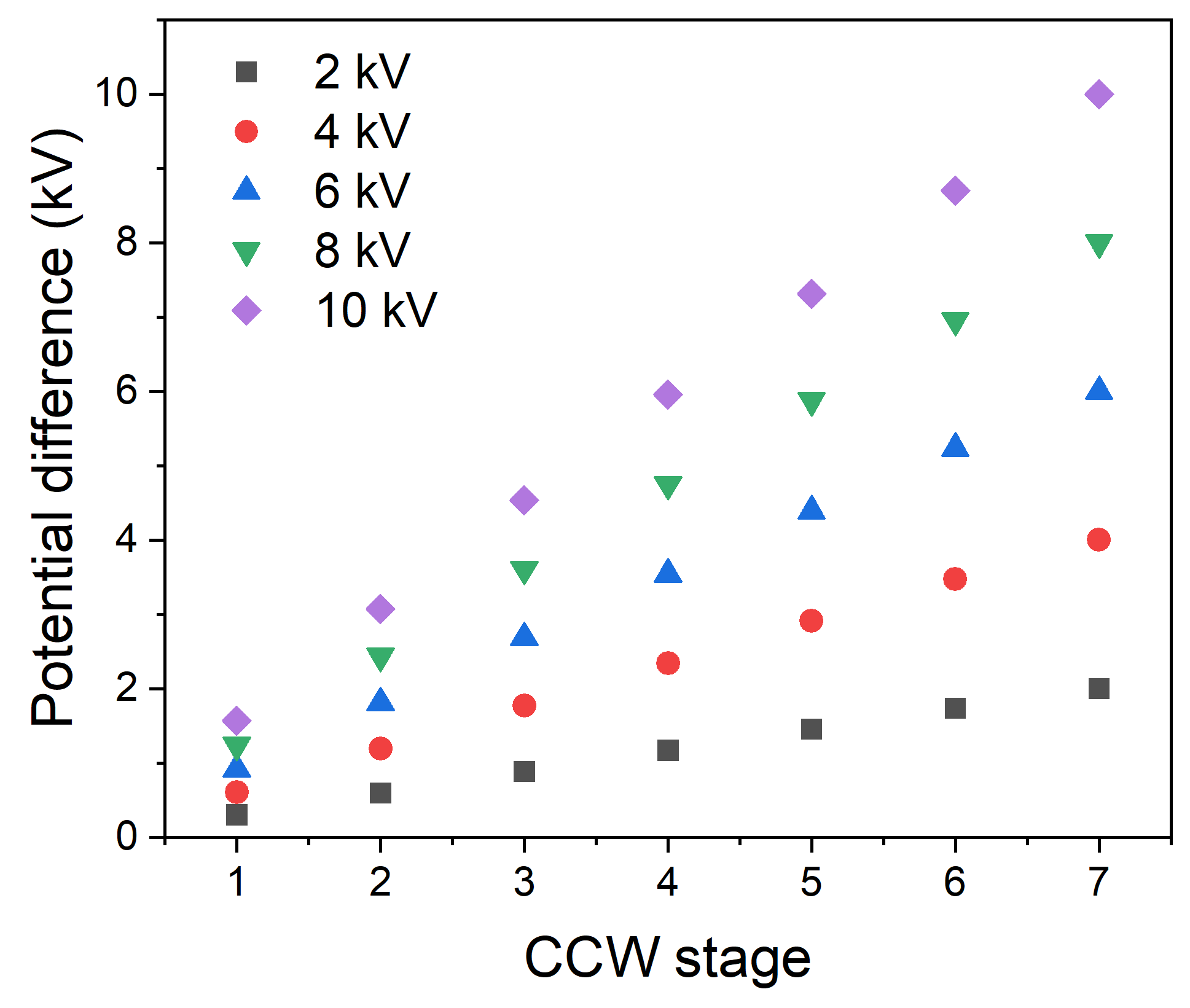}
    \caption{Measured voltage at successive stages of the CCW power supply for a given output voltage. 
    }
    \label{fig10}
\end{figure}

\begin{figure}[tbp]
    \includegraphics[width=0.48\textwidth]{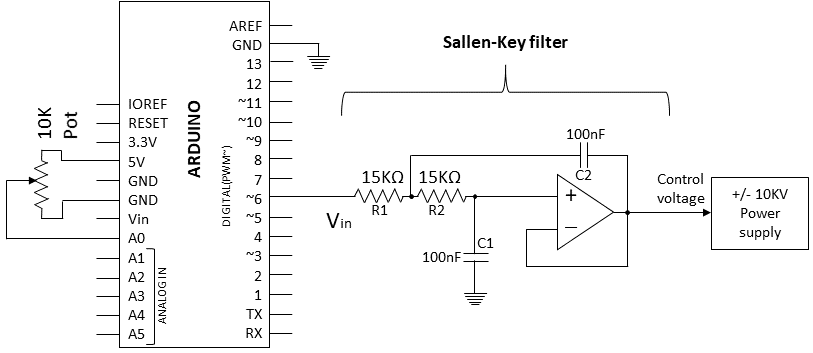}
    \caption{Schematic circuit diagram for remote operation of the power supply through Arduino micro controller and Sallen $-$ Key filter.}
    \label{fig11}
\end{figure}

\begin{figure*}[tbp]
    \centering
    \includegraphics[width=0.9\textwidth]{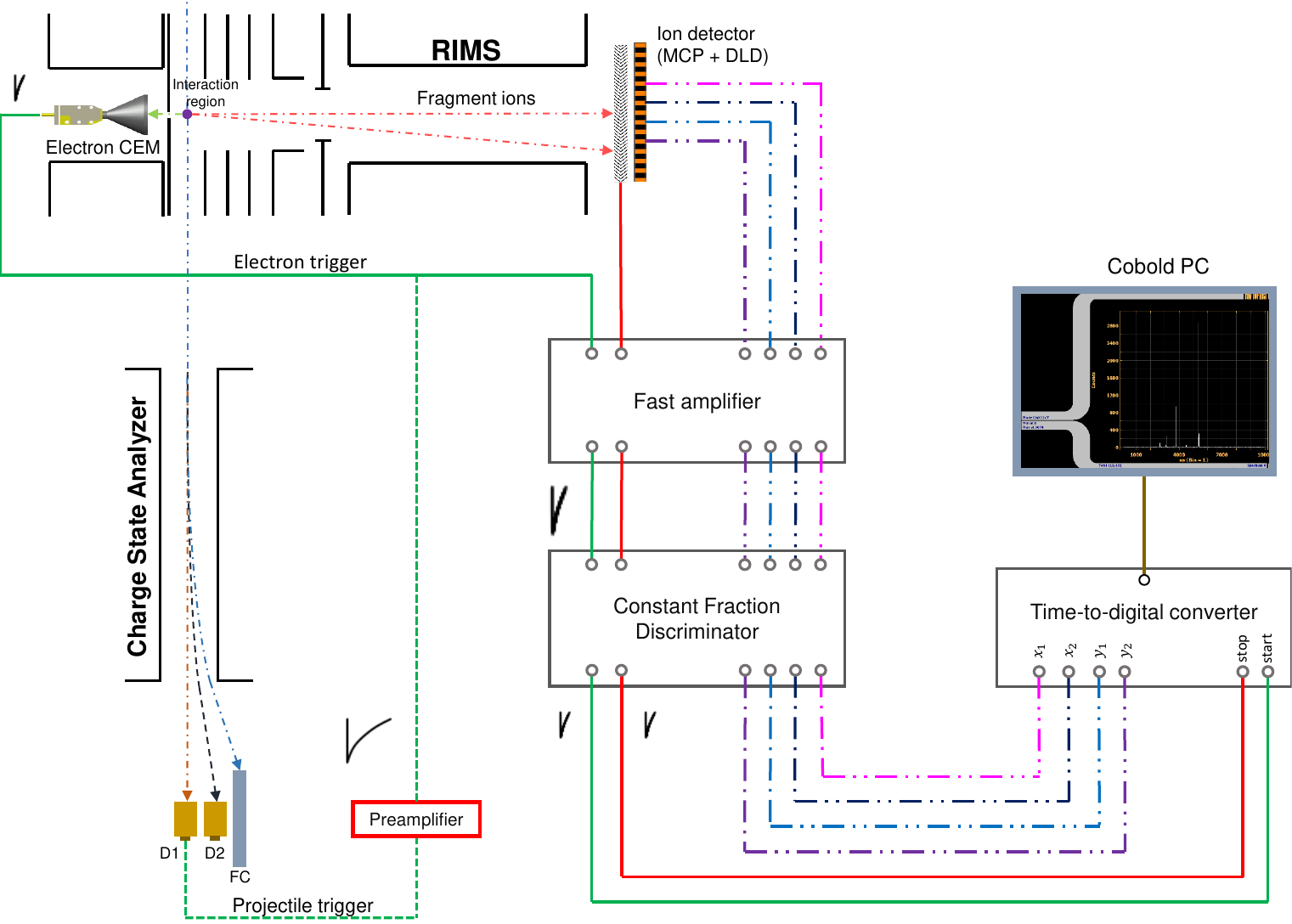}
    \caption{Schematic of the data acquisition system.}
    \label{fig12}
\end{figure*}

An oscillator circuit generates a 30 kHz square wave, which is subsequently amplified by a power amplifier consisting of a push-pull converter stage. The push-pull converter includes a step-up transformer with a primary-to-secondary turn ratio of 1:54. Here, the transformer is powered by a regulated 24 V dc power unit. The 24 V dc unit was also developed in-house using standard bridge rectifier circuit. The primary winding of the transformer is driven by two MOSFETs connected in a push-pull configuration, and are alternately switched ON and OFF by the oscillator signal. This switching action produces a high-frequency (30 kHz) square wave across the primary winding. The input and output waveforms of the transformer are shown in figure \ref{fig8}. The gate-source voltages $\mathrm{V_{GS1}}$ and $\mathrm{V_{GS2}}$ are input waveforms from the oscillator to the MOSFETs. The two waveforms are phase-shifted by $\pi$ radians. This phase-shift between the input and output waveforms results in generation of phase-shifted square waveforms $\mathrm{V_{DS1}}$ and $\mathrm{V_{DS2}}$ at the drains of the two MOSFETs $\mathrm{Q_1}$ and $\mathrm{Q_2}$. The final output generated between the terminals of primary winding of the transformer $\mathrm{T}$ is therefore a power amplified bipolar square wave of amplitude 24 V. The step-up action of the transformer yields an output voltage of 1.6 kV at the secondary winding. This 1.6 kV signal is then fed to a seven-stage CCW voltage multiplier circuit.

The CCW multiplier is designed to generate a bipolar output of $\pm$ 10 kV dc with a resolution of 11 V. The number of multiplier stages is determined based on (i) the required output voltage, (ii) the amplitude of the input voltage to the multiplier, and (iii) the voltage stress tolerance of the individual components. Notably, the output ripple voltage of the multiplier increases quadratically with the number of stages, imposing a trade-off between voltage gain and output quality. As a result, the final configuration represents an optimized balance among output voltage level, component ratings, physical size, and ripple characteristics. A control voltage is applied to a comparator circuit, which receives one of its inputs from the high-voltage output through a series resistor of 150 M$\Omega$. The comparator output regulates the primary side of the transformer, thereby controlling the secondary voltage that drives the CCW ladder. The control voltage varies linearly from 0 to 8.7 V corresponding to an output voltage range of 0 to 10 kV. This establishes a calibration ratio of 0.87 V per 1000 V (0.87:1000) for the control voltage, enabling accurate monitoring and regulation of the power supply output. In figure \ref{fig9} we have shown the calibration plot of control voltage vs high voltage output in no load and full load (200 M$\Omega$) conditions. As evident, the HV output varies linearly with the control voltage over the entire range of operation.

For a CCW voltage multiplier, the potential difference between successive stages is expected to be equal to the input voltage to the multiplier. However, in practical implementations, voltage drops occur across each stage due to the finite impedance of the capacitors. As a result, the potential difference between stages gradually decreases with increasing stage number. Figure \ref{fig10} shows the measured potential at various stages with respect to ground. The near-linear relationship observed between the stage number and potential indicates minimal losses and validates the high efficiency of the power supply. In order to evaluate the current handling capability of the power supply, a load resistor of 200 M$\Omega$ was connected across the output terminals at the maximum output voltage of 10 kV. Under this load, the power supply delivered a constant current of 50 $\mu$A over a continuous duration of two hours without any observable drop in output voltage. The 50 $\mu$A current sustenance of the power supply was tested over a range of output voltages.  Based on this performance, the upper limit on the current rating of the power supply was established at 50 $\mu$A. Further, the calculated load regulation is $2.3 \%$ which is well within the prescribed limits. The lower limit of the ac mains voltage for normal operation was determined by gradually reducing the input voltage using an autotransformer. It was observed that the power supply output began to drop below an input of 154 V (rms), establishing this value as the lower threshold for normal operation. Consequently, the ac main voltage range that ensures stable power supply operation was established as 154–230 V AC. To characterize the quality of the output of the power supply, we measured the ripple in its output voltage. The measurements were carried out using a high-voltage probe with an attenuation ratio of 1000:1, connected to a digital storage oscilloscope (UNI$-$T Model: UTD2102CEX+).  The recorded values have been summarized in table \ref{table4}. These ripple levels are within acceptable limits for the intended application, demonstrating the stability and effectiveness of the voltage multiplier design.

\begin{table}[tbp]
\centering
\begin{ruledtabular}
\begin{tabular}{ll}

Output Voltage (kV) & Ripple factor (in \%) \\

\hline

$+4.0 (-4.0)$ & 5.0 (5.0) \\

$+6.0 (-6.0)$ & 3.6 (3.3) \\

$+8.0 (-8.0)$ & 2.75 (2.5) \\

$+10.0 (-10.0)$ & 2.2 (2.5) \\

\end{tabular}
\end{ruledtabular}
\caption{Measured ripple voltage (in percent) across the output terminals of the power supply. The measurements were performed using a high-voltage probe of attenuation ratio 1000:1.}
\label{table4}
\end{table}

An Arduino Uno microcontroller based control module was developed to enable remote control of the power supply during collision experiments. The block diagram of the control module is shown in figure \ref{fig11}. In the remote control module, a variable analog dc signal is fed to the Arduino Uno microcontroller at one of the analog input pins (A0). The microcontroller is programmed to generate a PWM output corresponding to the input signal. The PWM signal is converted to a dc signal using a Sallen-Key filter \cite{}. The resulting dc voltage serves as the input control signal to the HV power supply. This implementation demonstrates the feasibility of using a microcontroller-based system for remote control of the power supply. In addition, one may also use wi-fi enabled microcontrollers for wireless operation of the CSA power supply. This is particularly useful in accelerator laboratories, where user access is restricted during the experiment.  

\begin{figure*}[tbp]
    \centering
    \includegraphics[width=\textwidth]{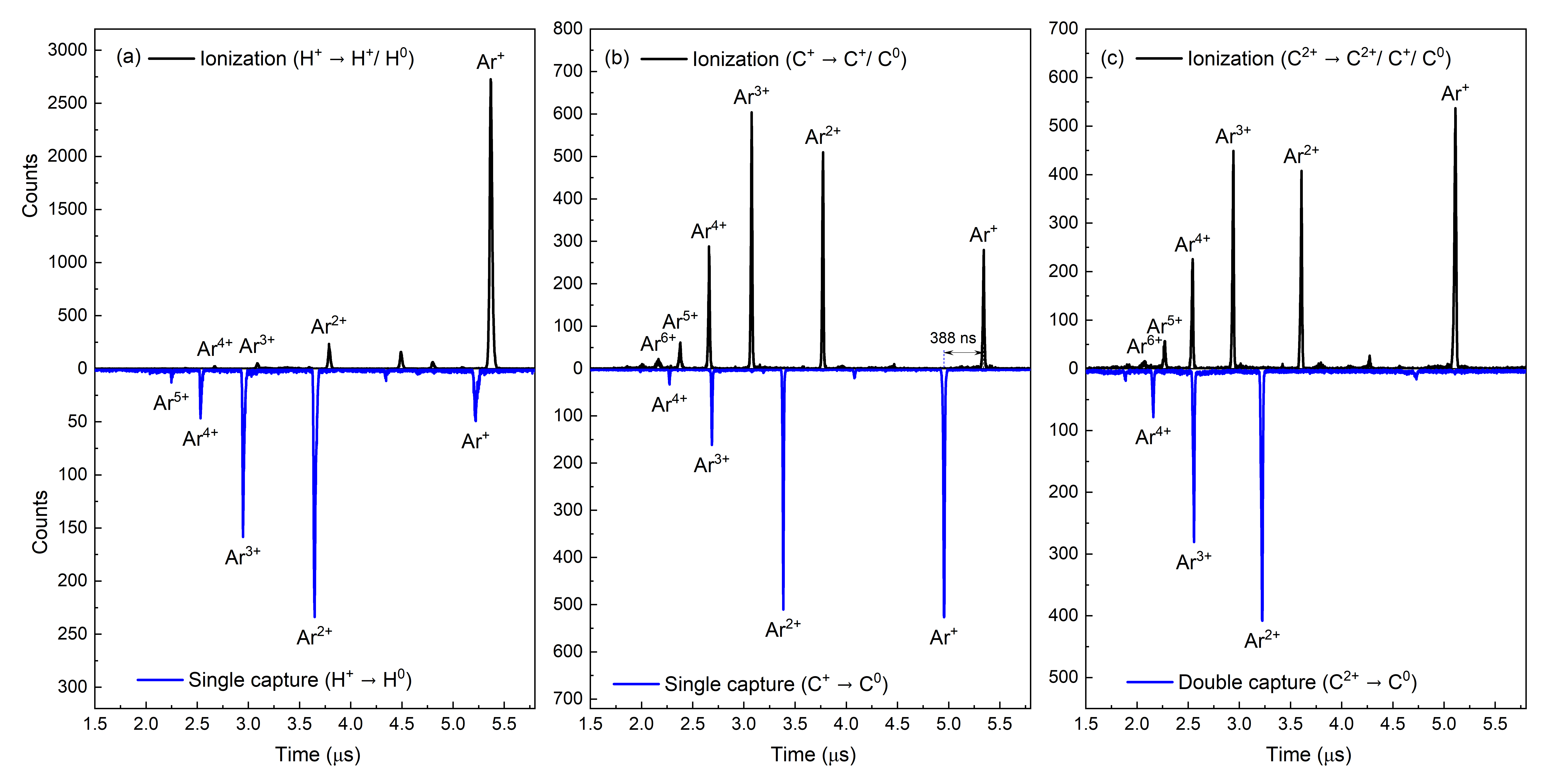}
    \caption{Time-of-flight spectrum (a) 1 MeV proton collision with Ar, (b) 1 MeV $\mathrm{C^+}$ collision with Ar and (c) 1 MeV $\mathrm{C^{2+}}$ collision with Ar. The upper panel is for ionization mode (electron trigger) and lower panel is for capture mode (neutral projectile trigger).}
    \label{fig13}
\end{figure*}

\begin{figure*}[tbp]
    \includegraphics[width = 0.9\textwidth]{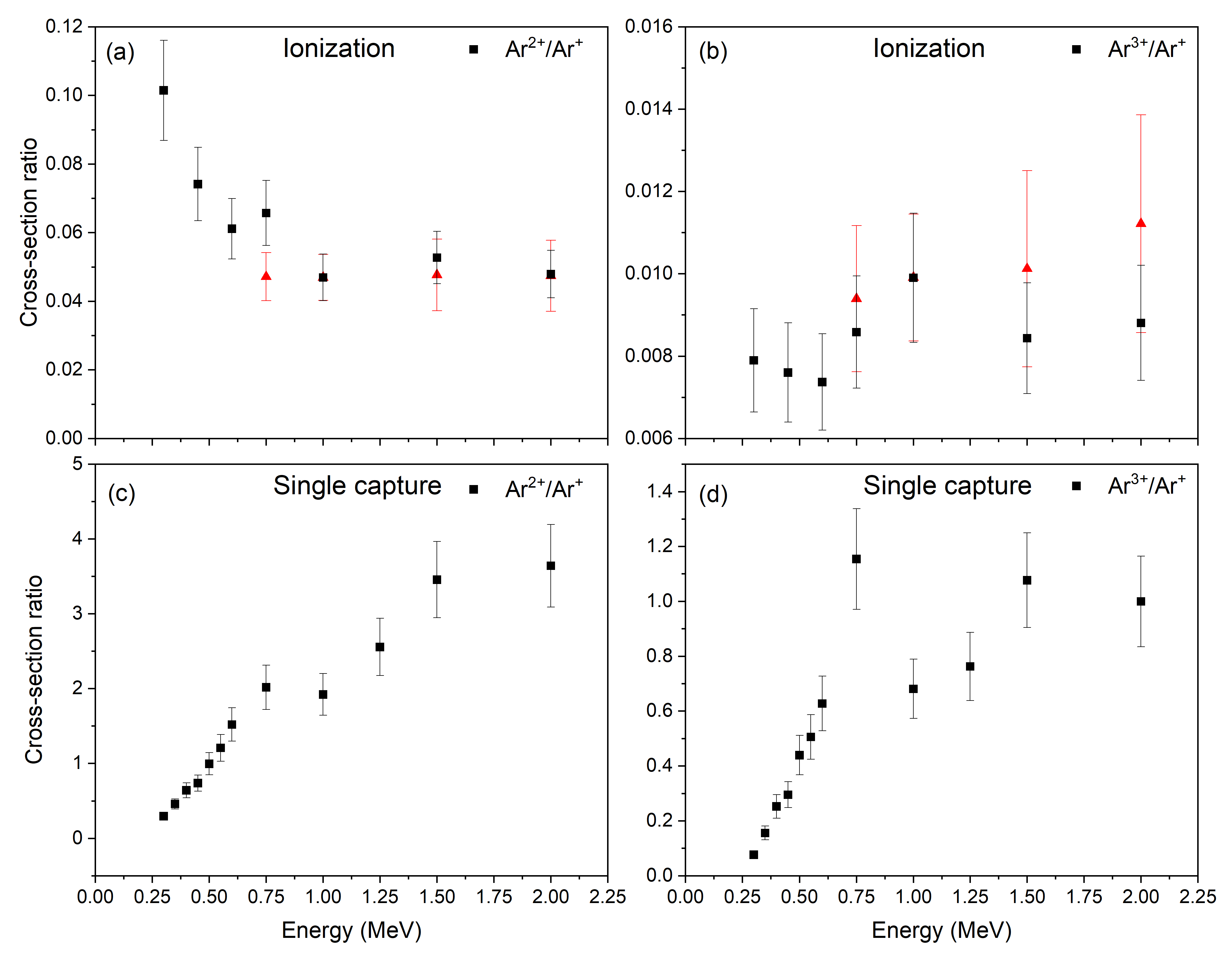}
    \caption{Projectile ($\mathrm{H^+}$) energy dependence of recoil ion yield ratio $\mathrm{Ar^{2+}/Ar^+}$ and $\mathrm{Ar^{3+}/Ar^+}$. Panel (a) and (b) show ionization mode (electron trigger). Panel (c) and (d) show capture mode (neutral projectile trigger). (Black) solid squares show data from the present measurement. (Red) solid triangle are ratios obtained from literature \cite{cavalcanti2003absolute}. The error bars represent the uncertainty in the reported absolute cross sections used for calculating the detector efficiency ratio \cite{cavalcanti2003absolute}. }
    \label{fig14}
\end{figure*}

\section{Collision experiments}

The collision experiments were performed at the 1.7 MV Tandetron accelerator facility at IIT Kanpur. Figure \ref{fig12} shows schematic of the data acquisition system. The RIMS setup has been discussed earlier in detail \cite{duley2022design}. In brief, projectile ions obtained from the accelerator collide with an effusive gas jet in the RIMS chamber. The overlap of the projectile and the target gas forms the interaction region. Electrons and the recoil ions generated in a collision event are extracted in opposite directions. The electrons are detected by a CEM detector at one end of the RIMS and the recoil ions are transported to the other end, where they hit a microchannel plate (MCP) with a delay line anode detector (DLD). The MCP + DLD assembly generates signals corresponding to the hit position (x,y) on the detector and TOF of the recoil ion. The CSA is connected to the RIMS scattering chamber. The projectile ions travel through CSA, post-collision, where they are charge state analyzed. The main projectile beam is detected on the FC whereas neutral projectile particles are detected by detector $\mathrm{D1}$  (see figure \ref{fig12}). In addition, for a doubly charged projectile beam, detector $\mathrm{D2}$ is employed to detect singly charged projectile ions (single capture processes).  Experiments were performed in two modes. (i) Ionization mode: In this mode signal from the electron detector was used as start signal for TOF measurement. (ii) Charge exchange mode: In this mode, the start signal for TOF measurements was taken from one of the projectile detectors, $\mathrm{D1}$ or $\mathrm{D2}$. Signals from all detectors were amplified using a fast amplifier and converted to a NIM standard timing signal using constant fraction discriminator before feeding to the multi-hit time-to-digital convertor (TDC). The signals from the TDC are processed in Cobold PC software \cite{} which stores the data event-by-event in list mode format for offline analysis.

\subsection{Ion $-$ atom collision}
We studied the ionization and electron capture yields for collision of MeV energy proton and carbon ions with Ar atomic target. In charged particle induced ionization of atomic targets, mainly three processes, pure ionization, pure capture, and transfer ionization, contribute to the total ionization yield. In our experimental
setup (RIMS + CSA), the data acquisition system works in twe modes, an electron "start" mode and (projectile) ion/neutral "start" mode. In the electron start mode the data are collected only for events corresponding to
ionization and transfer ionization processes, in which at least one electron is ejected to continuum after ion $-$ atom collision. In this mode, recoil ions created due to pure electron capture to the projectile are not recorded (due to absence of the electron start signal). Ionization events corresponding to pure capture and transfer ionization processes are recorded using the ion/neutral start mode. In this mode the start signal is taken from the ion/neutral detector of the CSA. In accelerator based experiments, the projectile ion beam travels a long distance pre $-$ collision. This results in charge state mixing in the primary projectile beam due to electron capture from background gases. This charge state mixing results in false coincidence recorded as capture events. The 1.7 MV tandetron accelerator beam line is equipped with a neutral beam trap \cite{duley2022design} to select only the primary charge state projectile before entering the collision chamber. Therefore the likelihood of false coincidences is minimized. During the experiment, the count rate of the ion/neutral detector in the CSA was kept at $\mathrm{\sim 100}$ Hz whereas, the background projectile detector count rate (target gas off) was observed to be $\mathrm{< 5}$ Hz, for the same incident projectile beam intensity. 
In figure \ref{fig13} we have shown the TOF spectra of Ar target corresponding to ionization (electron start mode) and capture (ion/neutral start mode) processes. The spectra are taken in collision with 1 MeV $\mathrm{H^+}$ (single capture), $\mathrm{C^+}$ (single capture) and $\mathrm{C^{2+}}$ (double capture) projectile ions. In all three cases, the neutral projectile atom signal was used as the start trigger. Up to six-fold ionization of Ar was observed in collisions with carbon ions. We can also see that the TOF of various $\mathrm{Ar^{q+}}$ ions corresponding to the capture process (lower panels) is shifted by $\mathrm{\sim 200}$ ns as compared to the ionization process (upper panels). This shift is attributed to the extra time that the projectile ion takes, after collision, to travel through the CSA. The time taken by the projectile to travel through the CSA depends on the velocity of the projectile beam.

In figure \ref{fig13}(a), we see that in the case of ionization (upper panel) the recoil ion yield is dominated by $\mathrm{Ar^+}$ ions and recoil ions up to $\mathrm{Ar^{4+}}$ are observed. In contrast, the lower panel in figure \ref{fig13}(a), corresponding to single capture events, shows a markedly different recoil ion distribution. The yield of $\mathrm{Ar^+}$ ion is considerably low compared to $\mathrm{Ar^{2+}}$ and $\mathrm{Ar^{3+}}$. We also see counts corresponding to $\mathrm{Ar^{5+}}$ recoil ion, which are almost negligible in the ionization channel. The higher recoil ion charge states in the capture channel correspond to transfer ionization process. The observed recoil ion charge state distribution shows that relative to pure capture process, transfer ionization is a dominant channel for $\mathrm{H^+}$ projectile in this collision energy range.

Figure \ref{fig13}(b) and (c) show the recoil ion yield distribution for ionization (upper panel) and capture (lower panel) channels for carbon projectile. As in the case of $\mathrm{H^+}$ projectile, the yield distribution of various recoil ions show marked differences between ionization and capture channels. The recoil ion yields corresponding to single and double capture have substantial contribution of higher charge states ($\mathrm{Ar^{3+}}$ and $\mathrm{Ar^{4+}}$) reflecting the importance of transfer ionization process. In figure \ref{fig13}(c) (lower panel), there is total absence of $\mathrm{Ar^+}$ recoil ion as expected for a double capture process.

In order to investigate the relative strengths of ionization, capture and transfer ionization processes, we have measured the recoil ion yields ($\mathrm{Y^q}$) of $\mathrm{Ar^{q+}}$ ions as a function of projectile energy for $\mathrm{H^+}$ projectile ion. The yield of a specific recoil ion is determined by integrating the area under the corresponding time-of-flight (ToF) peak and normalizing it with respect to the target gas density, projectile beam intensity and detector efficiency. A quantitative comparison between recoil ion yields from ionization and capture processes requires detector efficiency correction to this raw data. The present experimental setup is not suitable for determining the absolute detection efficiency of the MCP detector. This is on account of the unknown target gas density of the effusive gas jet in the interaction zone. Therefore, we have determined the yield ratio $\mathrm{R_q = Y_q/Y_1 (q > 1)}$. The yield ratio $\mathrm{R_q}$ is independent of the experimental parameters such as target gas density and projectile beam intensity. However the ratio depends on the absolute efficiency of the detector. We normalized our measured recoil ion yield ratios ($\mathrm{R_q}$) with the absolute total ionization cross section data reported by Cavalcanti et al. \cite{cavalcanti2003absolute} for 1 MeV proton impact on Ar, to correct for detector efficiency. In figure \ref{fig14}(a) and (b), we have shown the efficiency corrected recoil ion yield ratios, $\mathrm{R_2}$ and $\mathrm{R_3}$ as a function of proton beam energy for the ionization process. The absolute ionization cross section ratios derived from Cavalcanti et al. \cite{cavalcanti2003absolute} are also plotted (solid triangles). In both the plots (figure \ref{fig14}(a) and (b)), our measured ratios match well with those reported earlier. Using this detector efficiency correction factor, we have also plotted (see figure \ref{fig14}(c) and (d)) the yield ratios for the capture channel. The projectile energy was varied from 0.3 MeV ($\mathrm{v_p} = 3.4$ a.u.) to 2 MeV ($\mathrm{v_p} = 9$ a.u.). From figure \ref{fig14}(a) and (b) it is clear that, in this velocity range, single ionization dominates other higher ionization channels. As evident from the measured data, the double ionization cross section is more than an order of magnitude smaller that single ionization cross section. In addition, the double to single ionization cross section ratio decreases with increasing projectile velocity and falls by a factor of two up to 1 MeV, after which it seems to saturate. Similar observations are made for triple ionization channel (see figure \ref{fig14}(b)). The triple ionization cross section is almost two orders of magnitude smaller than single ionization cross section. However, the triple to single ionization cross section shows very weak dependence on the projectile velocity over the given energy range. The dependence of recoil ion ratios on the projectile energy for capture channel (see figure \ref{fig14}(c) and (d)) shows a notably different trend. In the electron capture mode (recoil ion coincidence with neutral H projectile), the recoil ion yield is dominated by single electron capture and transfer ionization processes. However, in this case we observe that the yield ratios $\mathrm{R_2}$ (figure \ref{fig14}(c)) and $\mathrm{R_3}$ (figure \ref{fig14}(d)) are of the order of unity. This implies that transfer ionization is more probable than pure single electron capture in this energy range. The yield ratios $\mathrm{R_2}$ and $\mathrm{R_3}$, both, increase monotonically with increasing $\mathrm{H^+}$ projectile energy. From figure \ref{fig14} (c) it is also evident that single electron capture dominates the recoil ion charge state distribution for proton energies below 0.5 MeV ($\mathrm{R_2} < 1$). At higher proton energies double ionization of Ar takes over single ionization ($\mathrm{R_2} > 1$). This behavior suggests a higher cross-section for transfer ionization process compared to single electron capture process. Similar trend is shown by the triple ionization channel where the yield ratio, $\mathrm{R_3}$ increases with increasing projectile energy and approaches unity for proton beam energy $\mathrm{> 1}$ MeV.

\subsection{Ion $-$ molecule collision}

\begin{figure*}[tbp]
    \includegraphics[width = \textwidth]{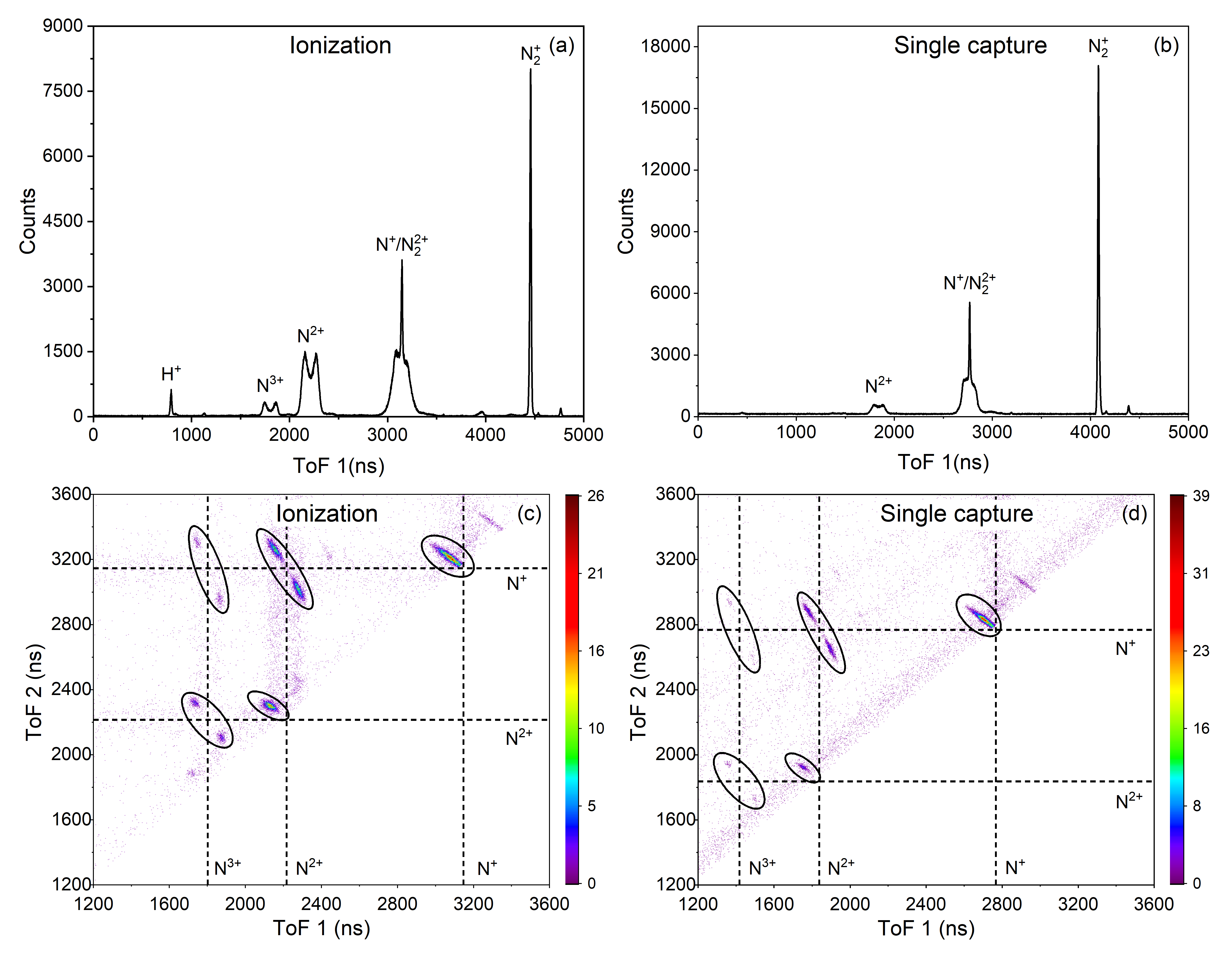}
    \caption{(a) Time-of-flight spectrum of $\mathrm{N_2}$ in collision with 1 MeV $\mathrm{C^+}$ projectile - Ionization mode (electron trigger). (b) Time-of-flight spectrum of $\mathrm{N_2}$ in collision with 1 MeV $\mathrm{C^+}$ projectile - Capture mode (neutral projectile trigger). (c) Ion $-$ ion coincidence diagram - Ionization mode. (d) Ion $-$ ion coincidence diagram - Capture mode. }
    \label{fig15}
\end{figure*}

\begin{figure*}[tbp]
    \includegraphics[width = \textwidth]{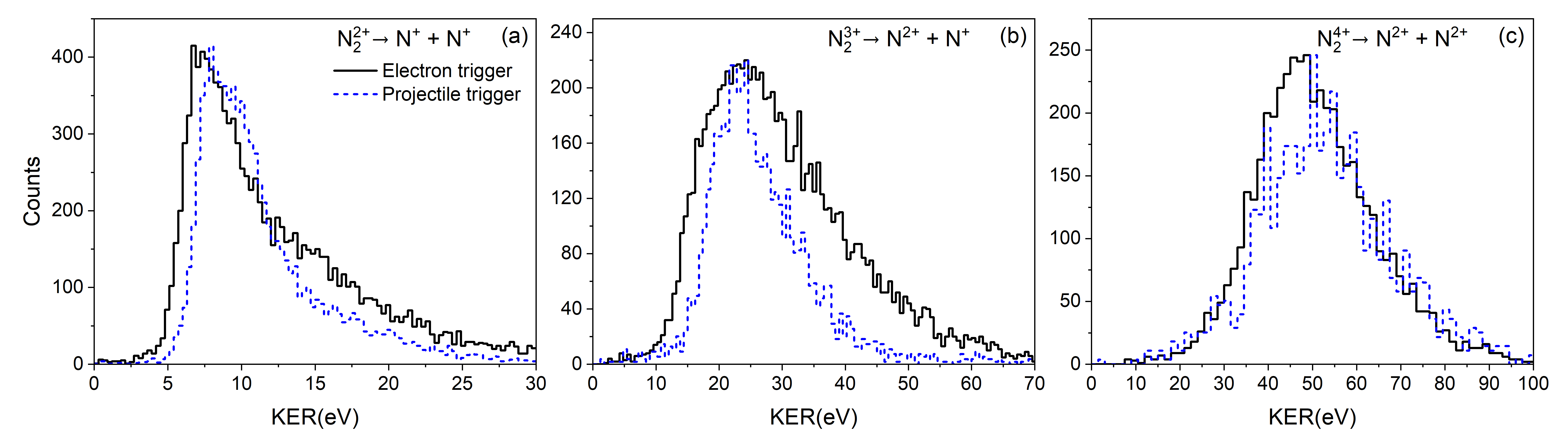}
    \caption{Kinetic energy release distribution for (a) $\mathrm{N_2^{2+} \longrightarrow N^+ + N^+}$ channel, (b) $\mathrm{N_2^{3+} \longrightarrow N^{2+} + N^+}$ channel and (c) $\mathrm{N_2^{4+} \longrightarrow N^{2+} + N^{2+}}$ channel. (Black) solid curve corresponds to ionization mode. (Blue) dashed curve corresponds to capture mode.}
    \label{fig16}
\end{figure*}

Ion collision induced multiple ionization and fragmentation of the N$_{\text{2}}$ molecule has been investigated in detail and could be used a model system to evaluate the performance of a the charge state analyzer. We studied the collision of ${\mathrm{N_2}}$ molecule with 1 MeV $\mathrm{C^+}$ ion beam in pure ionization mode ($\mathrm{e^-}$ signal as start trigger) and single electron capture mode (neutral $\mathrm{C^0}$ signal as start trigger). The TOF spectrum of ${\mathrm{N_2}}$ is shown in figure \ref{fig15}(a and b) where, atomic ion fragments up to $\mathrm{N^{3+}}$ are visible. For a homoatomic molecule such as N$_{\text{2}}$, the recoil ions N$_{\text{2}}^{\text{2+}}$ and N$^{\text{+}}$ have identical TOF due to identical mass to charge ratio. However, the two recoil ions, N$_{\text{2}}^{\text{2+}}$ and N$^{\text{+}}$, have different KERs. The molecular ion N$_{\text{2}}^{\text{2+}}$ is created in direct collision with the energetic projectile with negligible kinetic energy. Whereas the atomic ion N$^{\text{+}}$ is created via Coulomb fragmentation of N$_{\text{2}}^{\text{q+}}$ (q = 2,3) molecular ions and carries large kinetic energy. This difference in KER of the two ions is visible in the shape of the peak corresponding to N$_{\text{2}}^{\text{2+}}$/N$^{\text{+}}$ in the TOF spectrum. The narrow peak at the center is due to the parent N$_{\text{2}}^{\text{2+}}$ molecular ion, and the broader structure is due to the energetic N$^{\text{+}}$ ions. Also visible are the peaks corresponding to $\mathrm{N^{2+}}$ and $\mathrm{N^{3+}}$ atomic ions. These peaks result from Coulomb fragmentation of multiply charged $\mathrm{N_2^{q+}}$ (q = 3, 4, 5) molecular ions. In figure \ref{fig15}(a) (ionization mode), these peaks arise due to direct ionization and transfer ionization, whereas in figure \ref{fig15} (single capture mode) the atomic ion peaks are mainly due to the transfer ionization channel. This is also evident from the lower yield of atomic ions ($\mathrm{N^{q+}}$) compared to the singly ionized $\mathrm{N_2^+}$ molecular ions in the capture spectrum (figure \ref{fig15}(b)). The atomic ion peaks ($\mathrm{N^{2+}}$ and $\mathrm{N^{3+}}$) show a dip in the center, which is due to the loss of ions with larger kinetic energy ejected perpendicular to the extraction direction in the recoil ion momentum spectrometer. In addition, the yield of $\mathrm{N^{3+}}$ fragment is substantial in the ionization mode (figure \ref{fig15}(a)) whereas in the capture mode (figure \ref{fig15}(b)) this fragment is not visible. 

The fragmentation channels corresponding to various $\mathrm{N_2^{q+}}$ molecular ions can be distinguished in an ion-ion coincidence map or correlation diagram. In figure \ref{fig15} (c and d) we have shown the 2D TOF coincidence map for $\mathrm{N_2^{q+}}$ fragmentation. In the coincidence map, the TOF of the second hit (TOF2) is plotted against the TOF of the first hit (TOF1). Individual fragmentation channels show up as islands on such a map. Following fragmentation channels are observed in the present experiment.

\begin{equation}
    {\text{N}}_{\text{2}}^{\text{2+}} \longrightarrow {\text{N}}^{\text{+}} + {\text{N}}^{\text{+}}
\label{ch1}
\end{equation}
\begin{equation}
    {\text{N}}_{\text{2}}^{\text{3+}} \longrightarrow {\text{N}}^{\text{2+}} + {\text{N}}^{\text{+}}
\label{ch2}
\end{equation}
\begin{equation}
    {\text{N}}_{\text{2}}^{\text{4+}} \longrightarrow {\text{N}}^{\text{2+}} + {\text{N}}^{\text{2+}}
\label{ch1}
\end{equation}
\begin{equation}
    {\text{N}}_{\text{2}}^{\text{5+}} \longrightarrow {\text{N}}^{\text{3+}} + {\text{N}}^{\text{2+}}
\label{ch2}
\end{equation}

For two body fragmentation of homo$-$atomic molecular ions, momentum conservation, in the center-of-mass frame, governs the slope of the traces in the coincidence map given as \cite{duley2022design}:

\begin{equation}
    \mathrm{Slope = - \frac{q_{1}}{q_{2}}}
\end{equation}  

where $\mathrm{q_1}$ and $\mathrm{q_2}$ are the charge states of first and second ion, respectively. For equations (1) and (3)  above, the slope of the coincidence trace is $-1$ and for equation (2) and (4), the slopes are $-2$ and $-3$, respectively. 

\subsubsection*{Kinetic Energy Release Distribution}

The kinetic energy release distributions (KERD) for various fragmentation channel, equations (1), (2) and (3), are shown in figure \ref{fig16}(a) $-$ (c) for, both, ionization and capture processes. We have not shown the KERD corresponding to equation (4) due to very low statistics for this channel in the capture mode.

The KER spectrum corresponding to the $\mathrm{N_2^{2+} \longrightarrow N^+ + N^+}$ fragmentation channel (figure \ref{fig16}(a)) shows three energy components, at 7 eV, 9 eV and 14 eV. The KERD for ionization and capture modes is rather similar, with slight variation in the most probable energy values. The experimentally measured KER values are in excellent agreement with previously reported experimental and theoretical results \cite{hishikawa1998coulomb, khan2021velocity, siddiki2023role, wu2011coulomb, zhang2017dissociative, pandey2016probing, pandey2016electron, siegmann2000kinetic, rajput2006kinetic} ground state. The KER values in the high energy region are attributed to high lying repulsive states in the Frank-Condon region.

Figure \ref{fig16}(b) shows the KERD for $\mathrm{N_2^{3+} \longrightarrow N^{2+} + N^+}$ channel. The KERD corresponding to the ionization mode is broader than that corresponding to the capture mode. This may result from the population of low level excited states in case of capture induced excitation of the molecular ion. As remarked earlier, in the capture mode, multiply charged $\mathrm{N_2^{q+}}$ molecular ions are formed by the transfer ionization process and the measured KERD corresponds to the same. The most probable KER value measured for this channel is 21 eV, in agreement with previously reported data. This value is smallar than the maximum KER (25.4 eV) that can be released on dissociation of $\mathrm{N_2^{3+}}$ \cite{rajput2006kinetic}.

In figure \ref{fig16}(c), we have shown the KERD for  $\mathrm{N_2^{4+} \longrightarrow N^{2+} + N^{2+}}$ fragmentation channel. This is the charge symmetric dissociation channel for $\mathrm{N_2^{4+}}$ ion. The molecular ion may also dissociate via a charge asymmetric fragmentation channel in to  $\mathrm{N^{3+} + N^+}$ ions. This channel is visible in the ionization mode, however, it is not a prominent dissociation pathway in the capture mode (see figures \ref{fig15}(c and d)). Therefore, we have not shown the KERD corresponding to this channel. The most probable KER value for this channel is measured to be 46.5 eV (Ionization mode) and 49.5 eV (capture mode). The KER values can be attributed to (i) $\mathrm{N_2^{4+}(b ^5\Sigma) \longrightarrow N^{2+}(^2P) + N^{2+}(^4P)}$ and (ii) $\mathrm{N_2^{4+}(^3\Sigma) \longrightarrow N^{2+}(^2P) + N^{2+}(^2P)}$ dissociation pathways \cite{rajput2006kinetic}.

The asymptotic dissociation limits along with the molecular ion states, for these fragmentation channels, are summarized in table \ref{table5}
\renewcommand{\arraystretch}{1.3} 

\begin{table*}[tbp]
\small
\centering
\begin{ruledtabular}
\begin{tabular}{llllll}

Channel & Electronic states & Asymptotic limit & KER & KER & KER \\ 
 & & & (Present) & (Other) & (Theory) \\ 
 & & & [eV] & [eV] & [eV] \\ \hline

\ce{N2^{2+} -> N+ + N+} & A$^1\Pi_u$ & N$^+$($^3$P) + N$^+$($^3$P) & 7.11 (I) & 6.73--7 \cite{hishikawa1998coulomb, khan2021velocity, siddiki2023role, wu2011coulomb, zhang2017dissociative} & 6.743 \cite{lundqvist1996doppler} \\

& 1$^1\Sigma^+_u$ & N$^+$($^1$D) + N$^+$($^1$S) & & & 6.5--7.4 \cite{pandey2016probing} \\

& D$^3\Pi_g$ & N$^+$($^3$P) + N$^+$($^3$P) & 7.81 (C) & 7.4--7.5 \cite{pandey2016probing, khan2021velocity, rajput2006kinetic} & 7.532 \cite{lundqvist1996doppler} \\

& 1$^3\Sigma^+_u$ & N$^+$($^3$P) + N$^+$($^3$P) & & & 7.3--8.1 \cite{pandey2016probing} \\

& & & 9.14 (I) & 8.7--9.2 \cite{pandey2016probing, pandey2016electron, zhang2017dissociative} & \\

& 3$^1\Sigma_g^+$ & N$^+$($^1$D) + N$^+$($^1$D) & 9.7 (C) & 10.0 \cite{pandey2016electron, pandey2014charge, pandey2016probing} & 10.0 \cite{pandey2016probing} \\

& 1$^1\Sigma_u^-$ & N$^+$($^3$P) + N$^+$($^3$P) & 14.4 (I, C) & 14.5--15.6 \cite{siegmann2000kinetic, rajput2006kinetic, pandey2016electron} & 14.8 \cite{pandey2016probing} \\

\ce{N2^{3+} -> N^{2+} + N+} & $^2\Sigma$ & N$^{2+}$($^2$P) + N$^+(^1$S) & 21.7 (I), 21.5 (C) & 20 \cite{rajput2006kinetic} & 21.35 \cite{rajput2006kinetic} \\

& High-lying states & & 33.8 (I), 27.9 (C) & 29--32.3 \cite{siegmann2000kinetic, siddiki2023role, khan2021velocity} & \cite{siddiki2023role, khan2021velocity} \\

\ce{N2^{4+} -> N^{2+} + N^{2+}} & $^3\Sigma$ & N$^{2+}$($^2$P) + N$^{2+}$($^2$P) & 46.5 (I) & 42.8 \cite{siddiki2023role, rajput2006kinetic} & 46.59 \cite{rajput2006kinetic} \\

& & & 49.5 (C) & 51.3--54.9 \cite{siegmann2000kinetic, siddiki2023role} & \\

\end{tabular}
\end{ruledtabular}
\caption{Measured and calculated KER values for $\mathrm{N_2^{q+}}$ fragmentation channels with associated electronic state assignments and final product states in the asymptotic limit. I = Ionization, C = Capture.}
\label{table5}
\end{table*}

\section{\label{sec:level1}Conclusion\protect}
In conclusion, we have designed and successfully tested a projectile charge state analyzer (CSA) for electron capture-induced MeV energy ion–atom and ion–molecule collisions. The CSA is capable of detecting single and double electron capture events. The in-situ mechanism for switching between detectors expanded the analyzer’s applicability to both low- and high-energy collisions. A 10 kV bipolar power supply has also been developed, in-house, to operate the CSA. The CSA setup was augmented with the existing recoil ion momentum spectrometer to study capture induced atomic ionization and molecular fragmentation. Using the CSA, we investigated the dependence of ion yields $\mathrm{Ar^{q+}}$ on projectile energy by studying the collisions of Ar gas with proton ion beam. In addition, we have also studied the single electron capture-induced fragmentation of the \ce{N2} molecule. The measured kinetic energy release (KER) distributions for various fragmentation channels were found to be in good agreement with previous experimental and theoretical data. The present experimental setup could be used with anti$-$coincidence between the electron CEM signal and projectile detector signal to isolate contribution of pure ionization and pure single (or double) capture process. Furthermore, a three particle (electron $-$ recoil ion $-$ projectile ion (neutral)) coincidence will isolate the transfer ionization process.  

\begin{acknowledgments}
AHK would like to acknowledge the financial support received from Science and Engineering Research Board (SERB), Govt. of India via grant No. ECR/2017/002055.
\end{acknowledgments}

\section*{Author Contribution}
AHK and SBB designed the CSA. SBB carried out the simulations. SRS designed, assembled and tested the power supply. SBB, RD and RT performed the experiments. SBB and RD analyzed the data. SBB and AHK prepared the manuscript. 

\section*{Data availability statement}
The data that support the findings of this study are available from the corresponding author upon reasonable request.

\bibliography{apssamp}


\end{document}